\documentclass{PoS}
\usepackage{bm}% bold math
\newcommand{\bra}[1]{\langle \, #1 \, |}
\newcommand{\ket}[1]{| \, #1 \, \rangle}
\newcommand{\kket}[1]{\, #1 \, \rangle}
\newcommand{\physdim}[1]{\hspace{1ex} \mathrm{#1}}
\newcommand{\eqref}[1]{(\ref{#1})}

\title{Compositeness Of Quasibound States from Effective Field Theory}

\ShortTitle{Compositeness Of Quasibound States from Effective Field Theory}

\author{Yuki Kamiya\\
        Yukawa Institute for Theoretical Physics, Kyoto University, Kyoto 606-8502, Japan\\
        E-mail: \email{yuki.kamiya@yukawa.kyoto-u.ac.jp}}

\author{\speaker{Tetsuo Hyodo}\\
        Yukawa Institute for Theoretical Physics, Kyoto University, Kyoto 606-8502, Japan\\
        E-mail: \email{hyodo@yukawa.kyoto-u.ac.jp}}

\abstract{The internal structure of exotic hadron candidates are studied from the viewpoint of the compositeness of near-threshold states.
We focus on the Weinberg's weak-binding relation between the experimental observables and the compositeness of the state.
First we extend the relation for quasibound states within the framework of the effective field theory.
Next, considering the correction term, we develop a systematic method to estimate the uncertainty of compositeness in the weak-binding relation.
Finally, applying the extended relation and the error evaluation method, we conclude that the structure of $\Lambda(1405)$ is dominated by the $\bar{K}N$ component.}

\FullConference{The 26th International Nuclear Physics Conference\\
		11-16 September, 2016\\
		Adelaide, Australia}

\begin{document}

\section{Introduction}

With the discoveries of many candidates of the exotic hadrons~\cite{Olive:2016xmw,Brambilla:2010cs,Hosaka:2016pey}, which are not interpreted as the $qqq$ state nor  the $q\bar{q}$ state,
the investigation of the internal structure of these states is one of the most important topics in hadron physics.
To characterize the internal structure of these states, 
one of the useful methods is to evaluate the compositeness of the state, which is the probability of finding the composite component in the state, defined by the wave function~\cite{Weinberg:1965zz,Hyodo:2013nka,Sekihara:2014kya}.

Given  a certain model to calculate the wave function of the state or the scattering amplitude, the compositeness can be determined.
On the other hand, it is known that the compositeness $X$ of the weakly-bound state can be determined directly from a few experimental observables, thanks to the weak-binding relation:
\begin{equation}
	a_{0} = R\left\{ \frac{2X}{1+X} + {\mathcal O}\left(\frac{R_{\mathrm{typ}}}{R}\right)\right\},\quad R =1/\sqrt{2\mu B},
 \label{eq:comp-rel-bound}
\end{equation}
where $a_0$ is the scattering length, $B$ is the binding energy, $\mu$ is the reduced mass, $R$ is the length scale related to the binding energy and 
$R_{\mathrm{typ}}$ is the typical length scale of the interaction of the system. 
This relation implies that the leading term of the scattering length in the expansion of $R_{\mathrm{typ}}/R$ is determined only by the compositeness $X$. % 
In the weak-binding limit $R_{\mathrm{typ}}/R\ll 1$, 
the second term is negligible and the compositeness $X$ can be determined only with $a_0$ and $B$.
This weak-binding relation has been derived by Weinberg to show that the deuteron is a proton-neutron composite system~\cite{Weinberg:1965zz}.
The study of the compositeness of the physical hadrons is discussed in Refs.~\cite{Baru:2003qq,Aceti:2012dd,Hyodo:2013iga,Guo:2015daa}.

The direct determination of the compositeness with the weak-binding relation would be useful for exotic hadrons
because the sufficient amount of experimental data is not always accumulated.
In order to investigate the structure of exotic hadrons with finite decay width, the extension of the above relation to the quasibound state is necessary,
because the relation~\eqref{eq:comp-rel-bound} is valid only for the stable bound state. 
Furthermore, the higher order terms $\mathcal{O}(R_{\mathrm{typ}}/R)$ in Eq.~\eqref{eq:comp-rel-bound} is finite for the realistic hadron states and gives the finite correction to the compositeness.
To perform a systematic analysis, we need a method to evaluate the error of the compositeness in the weak-binding relation.
Here we present the extension of the weak-binding relation for the quasibound state based on the non-relativistic effective field theory and the method of the error evaluation of the compositeness. 
Finally, applying the extended weak-binding relation with the error evaluation to the $\Lambda(1405)$ baryon, we show the $\bar{K}N$ composite dominance in its structure.
All the contents in this paper is based on Refs.~\cite{Kamiya:2015aea,Kamiya:2016oao}.

\section{Extension of weak-binding relation to quasibound state}\label{sec:extention}
To generalize the weak-binding relation to the unstable quasibound states, 
let us consider the coupled-channel two-body s-wave scattering system. 
To treat this system, we introduce the non-relativistic effective field theory (EFT) with Hamiltonian
$H = H_{\mathrm{free}} + H_{\mathrm{int}}$~:
\begin{eqnarray}
H_{\mathrm
	{free}} &=&\int d^3\bm{r} \Biggl[\sum_{i=1,2}\left(\frac{1}{2 M_i} \mathbf{\nabla} \psi_i^\dagger \cdot\mathbf{\nabla} \psi_i +\frac{1}{2 m_i} \mathbf{\nabla} \phi_i^\dagger \cdot\mathbf{\nabla} \phi_i \right)
	\nonumber \\
	& & + \frac{1}{2M_{0}} \mathbf{\nabla}  B_0^\dagger \cdot{\mathbf \nabla} B_0 -\nu_\psi \psi_2^\dagger \psi_2-\nu_\phi \phi_2^\dagger \phi_2+\nu_0 B_0^\dagger B_0 \Biggr],\\
H_{\mathrm{int}} &=&\int d^3\bm{r} \left[\sum_{i=1,2} g_{0,i} \left( B_0^\dagger \phi_i\psi_i + \psi_i^\dagger\phi_i^\dagger B_0 \right) + \sum_{i,j=1,2} \lambda_{0,ij} \psi_j^\dagger\phi_j^\dagger \phi_i\psi_i\right],
\end{eqnarray}
with $\hbar=1$ and $\lambda_{0,12} = \lambda_{0,21}$. 
We consider that this EFT is applicable below the cutoff momentum $\Lambda$. 
The cutoff scale is related to the typical length scale of the interaction $R_{\mathrm{typ}}$ as $ \Lambda\sim 1/R_{\mathrm{typ}}$.
We set the threshold of the channel 2 ($\psi_2 \phi_2$) is lower than that of channel 1 ($\psi_1 \phi_1$).
To express the energy difference between the thresholds, we introduce $\nu \equiv  \nu_\psi + \nu_\phi > 0$. % 
We consider the system in which a quasibound state $\ket{QB}$ with eigenenergy $E_{QB}$ lies near the threshold energy of channel~1; $  H\ket{QB}=E_{QB}\ket{QB}$.
Due to the completeness relation which follows from the phase symmetry of the Hamiltonian, the physical quasibound state $\ket{QB}$ can be written as a linear combination of the scattering state and the discrete state as 
\begin{eqnarray}
  \ket{QB} 
  =c\ket{B_{0}}+\sum_{i=1,2}\int\frac{d^3\bm{p}}{(2\pi)^{3}} 
  \chi_i(\bm{p})\ket{\bm{p}_i}
  \label{eq:twobody2},
\end{eqnarray}
where $\ket{\bm{p}_i}=\tilde{\psi}_i^{\dag}(\bm{p})\tilde{\phi}_i^{\dag}(-\bm{p})/\sqrt{\mathcal{V}}\ket{0}$ and $\ket{B_{0}}=\tilde{B}_{0}^{\dag}(\bm{0})/\sqrt{\mathcal{V}}\ket{0}$ with the creation operators $\tilde{\psi}_i^{\dag}(\bm{p})$, the vacuum $\ket{0}$, and $\mathcal{V}=(2\pi)^{3}\delta^{3}(\bm{0})$.

The unstable state cannot be normalized with the ordinary normalization condition in the framework of quantum mechanics, $\langle QB|QB\rangle=1$.
To normalize the unstable state, we introduce the Gamow state $|\overline{QB} \rangle \equiv |QB \rangle ^*$. % old
With the convergence factor, one can show that the normalization condition $\langle \overline{QB}|QB\rangle = 1$ is well defined~\cite{Berggren:1968zz}. 
The compositeness and the elementariness are defined as % old
\begin{equation}
Z\equiv  \bra{\overline{QB}}\kket{B_0}\bra{B_{0}}\kket{QB}= c^2,
 \quad 
 X_i = \int \frac{d^3\bm{p}}{(2\pi)^{3}} \bra{\overline{QB}}\kket{\bm{p}_i}\bra{\bm{p}_i}\kket{QB}=\int \frac{d^3\bm{p}}{(2\pi)^{3}} \chi_i^2(\bm{p}),\label{eq:X-def-quasi} 
\end{equation}
which satisfy
\begin{equation}
Z+X_1+X_2=1,\quad Z,X_i \in \mathbb{C} .\label{eq:sum-rule-quasi} 
\end{equation}
The complex nature of $X_{i}$ and $Z$ reflects the fact that the expectation value of an arbitrary operator for  the unstable state becomes complex. Nevertheless, the sum rule~(\ref{eq:sum-rule-quasi}) is guaranteed owing to the normalization of $\ket{QB}$.

The forward scattering amplitude of the channel 1 is given by
\begin{eqnarray}
	f_{11}(E) =-\frac{\mu_{1}}{2\pi} \frac{1}{\left[v(E)\right]^{-1} - G_{1}(E) }, 
\end{eqnarray}
where $v(E)$ and $G(E)$ are defined as 
\begin{eqnarray}
v(E) &=& \lambda_{0,11}+\frac{g_{0,1}^{2}}{E-\nu_{0}}
   +\frac{\left(\lambda_{0,12}+\frac{g_{0,1}g_{0,2}}{E-\nu_{0}}\right)^{2}}
   {[G_2(E)]^{-1}-(\lambda_{0,22}+\frac{g_{0,2}^{2}}{E-\nu_{0}})} 
   \label{eq:vE},\\ 
G_i(E)
   &=&  
\int_{0}^{\Lambda}\frac{d^3\bm{p}}{(2\pi)^3} \frac{1}{E-p^2/(2\mu_i)+\delta_{i,2}\nu+i0^{+}},
\end{eqnarray}
with $\mu_i = m_iM_i/(m_i+M_i)$. 
By using these expressions, the compositeness of the channel 1, $X_1$, can be written as
\begin{eqnarray}
X_1 = \frac{1}{1 + G_{1}^{2}(E_{QB})v^\prime(E_{QB}) [G_1^\prime(E_{QB})]^{-1}}.
\end{eqnarray}

Now we redefine $R$ by $R \equiv 1/\sqrt{-2\mu_1 E_{QB}}$.
Expanding the scattering length $a_0$ in powers of $1/R$, 
we obtain the weak-binding relation for the quasibound state:
\begin{eqnarray}
a_0
=R \Biggl\{\frac{2X_1}{1+X_1} + {\mathcal O}\left(\left|\frac{R_{\mathrm{typ}}}{R}\right| \right) +  \mathcal{O} \left( \left| \frac{l}{R} \right|^{3}\right) \Biggr\}
\label{eq:comp-rel-quasi},
\end{eqnarray}
where $l=1/\sqrt{2\mu_{1}\nu}$ is defined with the energy difference between the thresholds $\nu$. 
Now we see that the first two terms are the same with Eq.~\eqref{eq:comp-rel-bound} and the contribution of the coupled channel appears as the third term in this extended weak-binding relation. %
The origin of this term is the expansion of the loop function of the decay channel $G_2(E)$. % %
This term is characterized with the additional length scale $l$. % 
When the energy difference $\nu$ is so small that $l$ satisfies $|l/R|^3 \ll 1$,
the higher order terms $\mathcal{O}(|l/R|^3)$ is negligible. 
In that case, Eq.~\eqref{eq:comp-rel-quasi} reduces to the same form with the weak-binding relation for the bound state~\eqref{eq:comp-rel-bound}. % % 

In the relation~\eqref{eq:comp-rel-quasi}, $a_0$ and $R$ are the model-independent physical quantities, which are the experimental observables.
On the other hand, the compositeness $X_1$ defined by the wave function of the state and the term of $\mathcal{O}(|R_{\mathrm{typ}}/R|)$ related to the detail of the interaction are  model-dependent.
When the length of the wave function $|R|$ is much larger than the scale of the interaction $R_{\mathrm{typ}}$, $\mathcal{O}(|R_{\mathrm{typ}}/R|)$ is small and can be neglected. 
By neglecting this term, the compositeness can be directly determined with model-independent quantities, $a_0$ and $R$.
This means that the model-dependence of the compositeness $X_1$ is also small in this case.
Thus we understand that only if the state is weakly bound, the compositeness can be model-independently determined by a few experimental observables. 

When $X_{1}$ is determined model independently, we can also determine $Z+X_{2}$ from Eq.~(\ref{eq:sum-rule-quasi}). 
In this case, however, separation of $Z$ and $X_{2}$ is model dependent. 
In the following, we rewrite $X_1$ as $X$ and $Z+X_2$ as $Z$. 
The sum rule is then given by $Z+X=1$ and $Z$ represents any contributions other than the channel 1.

\begin{figure}
\begin{center}
\includegraphics[width=7cm]{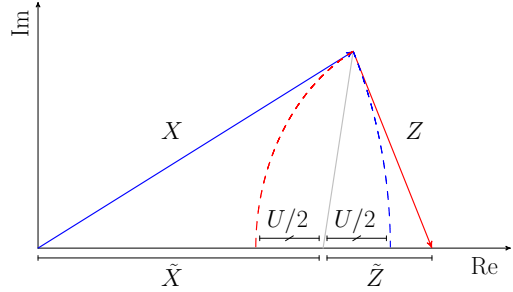}
\label{fig:Xtilde}
\caption{Geometric illustration of $\tilde{X}$, $\tilde{Z}$ and $U$ defined in Eq.~(2.12) %\eqref{eq:interpretation} 
quoted from Ref.~[12].}%\cite{Kamiya:2016oao}.}
\end{center}
\end{figure}

The compositeness of the quasibound state with the definition~(\ref{eq:X-def-quasi}) takes complex value, which cannot be interpreted as the probability. 
In Refs.~\cite{Kamiya:2015aea,Kamiya:2016oao}, we have introduced new quantities which allow us the probabilistic interpretation.
The real-valued compositeness (elementariness) $\tilde{X}$ ($\tilde{Z}$) and the uncertainty of the interpretation $U$ are defined by $X$ and $Z$ as
\begin{equation}
\tilde{X} \equiv \frac{1 - |Z| + |X|}{2},\quad \tilde{Z} \equiv \frac{1 - |X| + |Z|}{2},\quad 
    U \equiv |Z| +|X| -1 .\label{eq:interpretation} 
\end{equation} 
One can easily check that $\tilde{X}$ and $\tilde{Z}$ satisfy $\tilde{X} +\tilde{Z} =1 $ and $0\leq \tilde{X},\tilde{Z} \leq 1$, which are the necessary conditions to interpret $\tilde{X}$ and $\tilde{Z}$ as the probabilities.
A geometric illustration of this definition is given in Fig.~\ref{fig:Xtilde}.
In the bound state limit, $\mathrm{Im}\hspace{1ex} E_h=0$, $\tilde{X}$ and $\tilde{Z}$ equal to original $X$ and $Z$, $\tilde{X} =X, \tilde{Z} = Z$ and $U=0$. 
In this limit, $\tilde{X}$ and $\tilde{Z}$ are safely interpreted as the probabilities.
When the state has a small decay width, $X$ and $Z$ have a small finite imaginary part which makes the value of $U$ small but finite.
In this case, from Fig.~\ref{fig:Xtilde}, we can guess that $\tilde{X}$ and $\tilde{Z}$ acquire the uncertainty of $\pm U/2$. 
Thus, we regard the real-valued $\tilde{X}$ and $\tilde{Z}$ as the meaningful probabilities when the calculated $U/2$ is small enough.

\section{Estimation of the compositeness with error evaluation}\label{sec:error}
Here we consider the practical method to apply the weak-binding relation to the physical hadrons, when the higher order terms is small but not negligible. % %
Before discussing the quasibound state, let us consider the stable bound state.

In Eq.~\eqref{eq:comp-rel-bound}, the magnitude of the higher order terms of $\mathcal{O}(R_{\mathrm{typ}}/R)$ cannot be determined from the experimental observables because it is a model-dependent term. % %
However, for the shallow bound state with the small binding energy, this term should be irrelevant in Eq.~\eqref{eq:comp-rel-bound} and gives just a small correction to the estimation of the compositeness. % %
By neglecting this correction term, the central value of the compositeness $X_c$ can be determined by $a_0$ and $R$. % %
To evaluate the error which comes from the small correction term, we rewrite the relation as
\begin{eqnarray}
a_0 = R\left\{\frac{2X}{1+X} + \xi_r \right\},
\end{eqnarray}
where $\xi_r$ represents the higher order terms.
We consider to vary $\xi_r$ in the region to estimate the uncertainty of the compositeness~:
\begin{eqnarray}
   |\xi_{r}| & \leq R_{\mathrm{typ}}/R 
   \label{eq:condition_xir} .
\end{eqnarray}
One can determine $R_{\mathrm{typ}}$ by the lowest mass of the meson mediating the interaction as $R_{\mathrm{typ}} = 1/m_{\mathrm{typ}}$. % old ..
For the specific model calculation, $R_{\mathrm{typ}}$ can be  determined  with, for instance, by the spatial extent of the potential, or by the inverse of the momentum cutoff. % old 
With Eq.~\eqref{eq:condition_xir}, we obtain the central value and the upper and lower bounds of the compositeness as % old
\begin{eqnarray}
X_{c} = \frac{a_0/R}{2-a_0/R},\quad X_{u} = \frac{a_0/R+ \xi}{2-a_0/R - \xi},\quad X_l =  \frac{a_0/R- \xi}{2-a_0/R + \xi}, \quad\xi = R_{\mathrm{typ}}/R.\label{eq:ul_limit_X}
\end{eqnarray}
Namely, the uncertainty band of the compositeness is estimated by ${X_c}^{+(X_u - X_c)}_{-(X_c-X_l)}$. % old 

As an example, let us consider the compositeness of the deuteron.
The deuteron is a bound state in the $n$-$p$ scattering.
The typical length scale in the $n$-$p$ system is determined by the $\pi$ exchange interaction as $R_{\mathrm{typ}} = 1/ m_{\pi} = 1.43\physdim{fm}$. 
With the binding energy of the deuteron $B = 2.22\physdim{MeV}$ and the $n$-$p$ scattering length $a_0 =5.42\physdim{fm}$~\cite{Machleidt:2000ge},
the $n$-$p$ compositeness $X$ is estimated as $ X = 1.68^{+2.15}_{-0.83}.$
We note that the central value $X_c$ is greater than unity, which cannot be interpreted as the probability.
However, the uncertainty band includes the $X\sim1$ region.
The large uncertainty band is caused by $\xi = 0.331$, which is not very small even for the  binding energy $2.22\physdim{MeV}$ due to the $\pi$ exchange. 
In this way, the adequate treatment of the correction term confirms the $n$-$p$ composite nature of the deuteron.

Now let us consider the estimation of the compositeness of the quasibound state.
Again the central value of the compositeness $X$ and $\tilde{X}$ are determined from the scattering length $a_0$ and the eigenenergy $E_{QB}$ by neglecting the correction terms in Eq.~\eqref{eq:comp-rel-quasi}. % %
When we consider the contribution of the correction term to the compositeness $X$, 
we evaluate the uncertainty band of the real-valued $\tilde{X}$, for which the probabilistic interpretation is applicable.
To estimate the effect of the higher order terms, we first introduce a complex quantity $\xi_{c}$ in the expression of the weak-binding relation as
\begin{eqnarray}
a_0 = R\left\{\frac{2X}{1+X} + \xi_c \right\}.
\end{eqnarray}
Then the compositeness is obtained by
\begin{eqnarray}
X = \frac{a_0/R + \xi_{c}}{2-a_0/R-\xi_{c}}\label{eq:estimate_error_quasi}.
\end{eqnarray}
In the present case, $\xi_{c}$ is made of two components $\mathcal{O}(|R_{\mathrm{typ}}/R|)$ and $\mathcal{O}(|l/R|^3)$.
Both terms are  in general complex with an unknown relative phase. 
As a conservative error estimation, we allow $\xi_{c}$ to vary in the region
\begin{eqnarray}
   |\xi_{c}| & \leq |R_{\mathrm{typ}}/R|+|l/R|^3 
   \label{eq:condition} .
\end{eqnarray}
In other words, the largest magnitude of $\xi_{c}$ is determined when two terms are coherently added. We then evaluate $\tilde{X}$ with Eq.~\eqref{eq:interpretation} by varying $\xi_{c}$ with Eq.~\eqref{eq:condition} being the constraint. Denoting the maximum (minimum) value of $\tilde{X}$ as $\tilde{X}_{u}$ ($\tilde{X}_{l}$), we consider the uncertainty band of $\tilde{X}$ as $\tilde{X}_u<\tilde{X} < \tilde{X}_l$.

\section{Application to $\Lambda(1405)$}\label{sec:Application}
Here we apply the extended weak-binding relation with the error analysis to the $\Lambda(1405)$ baryon to discuss its internal structure.
The $\Lambda(1405)$ resonance with $J^{P} = 1/2^{-}$ lies near the $\bar{K}N$ threshold energy~\cite{Olive:2016xmw}.
By the resent studies with the chiral SU(3) dynamics, it is pointed out that this resonance is associated with two poles in the complex energy plane~\cite{Ikeda:2011pi,Ikeda:2012au,Mai:2012dt,Guo:2012vv,Mai:2014xna}.
In the present study, we focus on the $\bar{K}N$ compositeness of the pole which is closer to the $\bar{K}N$ threshold energy.

With the chiral SU(3) dynamics, the eigenenergy of the pole $E_h$ and the $I=0$ $\bar{K}N$ scattering length $a_0$ are obtained in the experimental  analyses of  Refs.~\cite{Ikeda:2011pi,Ikeda:2012au,Mai:2012dt,Guo:2012vv,Mai:2014xna} as shown in Table~\ref{tab:Lambda}.
The parameter sets 1, 2, 3, 4 and 5 are based on the NLO amplitude of Refs.~\cite{Ikeda:2011pi,Ikeda:2012au}, Ref.~\cite{Mai:2012dt}, Fit II of Ref.~\cite{Guo:2012vv}, solutions \# 2 and \# 4 of Ref.~\cite{Mai:2014xna}, respectively.
Because the values of $E_h$ and $a_0$ are scattered among the analyses, 
we estimate the systematic error by comparing all of these results.

Before estimating the compositeness, let us check the magnitude of the higher order terms.
We estimate the  interaction range $R_{\mathrm{typ}}$ from the $\rho$ meson exchange as $R_{\mathrm{typ}} =1/m_\rho \sim 0.25 \physdim{fm}$, because the  pion exchange interaction is prohibited between $\bar{K}$ and $N$. 
The length scale $l=1/\sqrt{2\mu\nu}$ can be determined by the 
the difference between the threshold energies of the $\bar{K}N$ channel and the $\pi\Sigma$ channel, $\nu = 104\physdim{MeV}$. 
Because $|R_{\mathrm{typ}}/R| \lesssim0.17$ and $|l/R|^3\lesssim0.14$ are not so large, 
the estimation of the compositeness with the weak-binding relation is applicable to $\Lambda(1405)$.

Now applying the method constructed in Secs.~\ref{sec:extention} and \ref{sec:error}, we calculate the compositeness $X_{\bar{K}N}$ and $\tilde{X}_{\bar{K}N}$ with the error analysis, and the uncertainty of the interpretation $U/2$ as summarized in Table~\ref{tab:Lambda}.
The results of $U/2$ show the small uncertainty of the interpretation of $\tilde{X}_{\bar{K}N}$, which comes from the unstable property of $\Lambda(1405)$.
Thus we can interpret the  value of $\tilde{X}_{\bar{K}N}$ as the probability.

From the results of $\tilde{X}_{\bar{K}N}$ with the error analysis, the central values of $\tilde{X}_{\bar{K}N}$ in all cases are consistently close to unity.
With the uncertainty band, the value of the $\tilde{X}_{\bar{K}N}$ is larger than $1/2$.
Thus we conclude that $\Lambda(1405)$ is dominated by the $\bar{K}N$ composite component.

\begin{table}[bt]
		\caption{Properties and results for the higher energy pole of $\Lambda (1405)$. Shown are the eigenenergy $E_{h}$, the $\bar{K}N(I=0)$ scattering length $a_{0}$, the magnitude of higher order terms $|R_{\mathrm{typ}}/R|$ and $|l/R|^3$, the $\bar{K}N$ compositeness $X_{\bar{K}N}$ and $\tilde{X}_{\bar{K}N}$ and the uncertainty of the interpretation $U/2$.}
		\label{tab:Lambda}
	\begin{center}
		\begin{tabular}{lccccccc}
			\hline
              & $E_{h}\physdim{[MeV]}$ & $a_0 \physdim{[fm]} $ &    $|R_{\mathrm{typ}}/R|$ & $|l/R|^3$ & $X_{\bar{K}N}$ & $\tilde{X}_{\bar{K}N}$& $U/2$   \\  \hline
			 Set 1 \cite{Ikeda:2012au}  & $-10-i26$ & $1.39 - i 0.85$ 
			 & $0.17$ & $0.14$ &  $1.2+i0.1$ &$1.0^{+0.0}_{-0.4}$ & $0.3$  \\ 
			 Set 2 \cite{Mai:2012dt}  & $-\phantom{0}4-i\phantom{0}8$ & $1.81-i0.92$ 
			 & $0.10$ & $0.03$ &$0.6+i0.1$ &  $0.6^{+0.2}_{-0.1}$ & $0.0$ \\ 
			 Set 3 \cite{Guo:2012vv}  & $-13-i20$ & $1.30-i0.85$ 
			 &  $0.16$ & $0.11$   & $0.9-i0.2$ &$0.9^{+0.1}_{-0.4}$ & $0.1$  \\
			 Set 4 \cite{Mai:2014xna}  & $\phantom{-0}2-i10$ & $1.21-i1.47$ 
			 & $0.10$ & $0.03$ &$0.6+i0.0$ &  $0.6^{+0.3}_{-0.1}$ & $0.0$ \\ 
			 Set 5 \cite{Mai:2014xna}  & $-\phantom{0}3-i12$ & $1.52-i1.85$ 
			 &  $0.12$ & $0.04$ & $1.0+i0.5$ &$0.8^{+0.2}_{-0.2}$ & $0.3$ \\ 
			 \hline
		\end{tabular} 
	\end{center}
\end{table}

\section{Conclusion}\label{sec:conclusion}
We have discussed the application of the weak-binding relation to the physical hadron state.
We have presented the derivation of the extended weak-binding relation of the quasibound state based on the non-relativistic effective field theory.
The relation tells us that the compositeness can be model-independently determined with the experimental observables in the weak-binding limit.
The interpretation of the complex value of the compositeness is discussed by introducing the new quantity $\tilde{X}$.
We also discuss the contribution of the higher order terms in the weak-binding relation.
The way to estimate the error of the compositeness by the higher order terms is constructed.
Finally, using the extended relation with the error analysis, we find that the $\bar{K}N$ composite component is dominant in $\Lambda(1405)$.
We consider that the present approach facilitates the future investigation of the hadron structure.

%
%\bibliographystyle{h-physrev3}
%%\bibliographystyle{plain}
%\bibliography{../../../bibtex/ref} 

\end{document}